\documentclass[twoside,11pt]{article}

\usepackage[preprint]{jmlr2e}
\usepackage[utf8]{inputenc}
\usepackage[T1]{fontenc}
\usepackage{url}
\usepackage{booktabs}
\usepackage{amsfonts}
\usepackage{amsmath}
\usepackage{amsopn}
\usepackage{enumitem}
\usepackage{pifont}
\usepackage{nicefrac}
\usepackage{microtype}
\usepackage[dvipsnames]{xcolor}
\usepackage{lipsum}
\usepackage{cleveref}
\usepackage{kmath}
\usepackage{glossaries}
\usepackage{xspace}
\usepackage{graphicx}
\usepackage{subcaption}
\usepackage{tikz}
\usepackage{wrapfig}
\usetikzlibrary{calc,shapes,arrows,positioning}

\usepackage{listings}
\usepackage{xcolor}
\usepackage{lastpage}
\usepackage{siunitx}

\definecolor{codebg}{RGB}{248,248,248}
\definecolor{keywordcolor}{RGB}{32, 74, 135}
\definecolor{stringcolor}{RGB}{78, 154, 6}
\definecolor{commentcolor}{RGB}{143, 89, 2}
\definecolor{classcolor}{RGB}{128, 0, 128}
\definecolor{functioncolor}{RGB}{0, 102, 204}
\definecolor{operatorcolor}{RGB}{206, 92, 0}
\definecolor{identifiercolor}{RGB}{0, 0, 0}
\definecolor{numbercolor}{RGB}{0, 0, 180}

\lstdefinestyle{mystyle}{
    language=Python,
    backgroundcolor=\color{codebg},
    basicstyle=\footnotesize\tt,
    morekeywords={as,self,float,int},
    keywordstyle=\color{keywordcolor},
    stringstyle=\color{stringcolor},
    commentstyle=\color{commentcolor},
    identifierstyle=\color{identifiercolor},
    literate=
        *{=}{{{\color{operatorcolor}=}}}{1}
        {+}{{{\color{operatorcolor}+}}}{1}
        {-}{{{\color{operatorcolor}-}}}{1}
        {*}{{{\color{operatorcolor}*}}}{1}
        {/}{{{\color{operatorcolor}/}}}{1}
        {==}{{{\color{operatorcolor}==}}}{2}
        {!=}{{{\color{operatorcolor}\neq}}}{2}
        {<=}{{{\color{operatorcolor}\leq}}}{2}
        {>=}{{{\color{operatorcolor}\geq}}}{2}
        {<}{{{\color{operatorcolor}<}}}{1}
        {>}{{{\color{operatorcolor}>}}}{1}
        {+=}{{{\color{operatorcolor}+=}}}{2}
        {<=}{{{\color{operatorcolor}<=}}}{2}
        ,
    morekeywords=[2]{abs},
    keywordstyle=[2]\color{identifiercolor}\bfseries,
    showstringspaces=false,
    tabsize=4,
    breaklines=true,
    frame=lines,
    framerule=1pt,
    rulecolor=\color{gray},
    framexleftmargin=1em,
    framexrightmargin=1em,
}
 
\newcommand{\ppb}{P}

\newcommand{\pobj}{p}
\newcommand{\lbpobj}{\LB{p}}

\newcommand{\pvletter}{x}
\newcommand{\bvletter}{z}
\newcommand{\dvletter}{u}

\newcommand{\wvletter}{w}
\newcommand{\pv}{\mathbf{\pvletter}}

\newcommand{\dv}{\mathbf{\dvletter}}

\newcommand{\wv}{\mathbf{\wvletter}}
\newcommand{\pvi}[1]{\pvletter_{#1}}
\newcommand{\bvi}[1]{\bvletter_{#1}}

\newcommand{\wvi}[1]{\wvletter_{#1}}

\newcommand{\LB}[1]{\tilde{#1}}
\newcommand{\UB}[1]{\bar{#1}}

\newcommand{\pdim}{n}
\newcommand{\ddim}{m}
\newcommand{\reg}{\lambda}
\newcommand{\lfunc}{f}
\newcommand{\pfunc}{h}
\newcommand{\rfunc}{g}

\newcommand{\dfunc}{\LB{D}}

\newcommand{\bigM}{M}
\newcommand{\regone}{\alpha}
\newcommand{\regtwo}{\beta}
\newcommand{\dic}{\mathbf{A}}
\newcommand{\dici}[1]{\mathbf{a}_{#1}}
\newcommand{\atom}[1]{\mathbf{a}_{#1}}
\newcommand{\obs}{\mathbf{y}}
\newcommand{\obsi}[1]{y_{#1}}

\newcommand{\lipschitz}{L}

\newcommand{\idxentry}{i}
\newcommand{\nodesymbol}{\nu}
\newcommand{\node}[1]{#1^\nu}

\newcommand{\setzero}{\nodesymbol_0}
\newcommand{\setone}{\nodesymbol_1}

\newcommand{\subzero}[1]{#1_{\setzero}}
\newcommand{\subone}[1]{#1_{\setone}}

\newcommand{\childnode}[3]{#1_{#2,#3}}
\newcommand{\pertslope}{\tau}
\newcommand{\pertlimit}{\mu}
\newcommand{\pertrightlope}{\kappa}
\newcommand{\relaxregfunc}{\LB{\rfunc}}

\newcommand{\workingset}{\mathcal{W}}
\newcommand{\violationset}{\mathcal{V}}
\newcommand{\spherecenter}{\mathbf{c}}
\newcommand{\sphereradius}{r}

\newcommand{\elops}{\texttt{El0ps}\xspace}
\newcommand{\scikit}{\texttt{Scikit-learn}\xspace}
\newcommand{\lobnb}{\texttt{L0bnb}\xspace}
\newcommand{\mimosa}{\texttt{Mimosa}\xspace}

\newcommand{\gurobi}{\texttt{Gurobi}\xspace}

\newcommand{\pybnb}{\texttt{Pybnb}\xspace}

\newcommand{\skglm}{\texttt{Skglm}\xspace}

\newcommand{\numpy}{\texttt{Numpy}\xspace}
\newcommand{\numba}{\texttt{Numba}\xspace}
\newcommand{\oa}{\texttt{Oa}\xspace}

\DeclareMathOperator{\argmin}{argmin}
\DeclareMathOperator{\argmax}{argmax}

\DeclareMathOperator{\dom}{dom}

\DeclareMathOperator{\gap}{gap}
\DeclareMathOperator{\interior}{int}
\DeclareMathOperator{\prox}{prox}

\newcommand{\0}{\mathbf{0}}
\newcommand{\abs}[1]{|#1|}

\newcommand{\conj}[1]{#1^{\star}}
\newcommand{\biconj}[1]{#1^{\star\star}}
\newcommand{\grad}{\nabla}
\newcommand{\icvx}{\iota}
\newcommand{\intervint}[2]{\{#1,\dots,#2\}}

\newcommand{\norm}[2]{\|#1\|_#2}
\newcommand{\opt}[1]{#1^{\star}}
\newcommand{\pospart}[1]{[#1]_+}
\newcommand{\separable}[2]{#1_{#2}}
\newcommand{\subdiff}{\partial}
\newcommand{\transp}[1]{#1^{\mathrm{T}}}

\newcommand{\eg}{\textit{e.g.}}

\if@final
  \newcommand{\AddNote}[1]{}
  \newcommand{\AddTodo}[1]{}

  \newcommand{\ProvideEditionMacros}[2]{%
    \expandafter\NewDocumentCommand\csname Add#1\endcsname{gg}{}%
    \expandafter\NewDocumentCommand\csname Rem#1\endcsname{gg}{}%
    \expandafter\NewDocumentCommand\csname Sup#1\endcsname{gg}{}%
  }
\else
  \newcommand{\AddNote}[1]{\textcolor{red}{[Note: #1]}}
  \newcommand{\AddTodo}[1]{\textcolor{red}{[Todo: #1]}}

  \newcommand{\ProvideEditionMacros}[2]{%
    \expandafter\NewDocumentCommand\csname Add#1\endcsname{gg}{\textcolor{#2}{##1}}%
    \expandafter\NewDocumentCommand\csname Rem#1\endcsname{gg}{\textcolor{#2}{[#1: ##1]}}%
    \expandafter\NewDocumentCommand\csname Sup#1\endcsname{gg}{\textcolor{#2}{\st{##1}}}%
  }
\fi

\newacronym{bfs}{BFS}{Best-First Search}
\newacronym{bnb}{BnB}{Branch-and-Bound}
\newacronym{mip}{MIP}{Mixed-Integer Programming}
\newacronym{oa}{OA}{Outer-Approximation}

\ProvideEditionMacros{CE}{purple}
\ProvideEditionMacros{CH}{blue}
\ProvideEditionMacros{TG}{orange}

\newcommand{\codewidth}{0.9}

\jmlrheading{XX}{2025}{1-\pageref{LastPage}}{XX/XX; Revised XX/XX}{XX/XX}{XX-XXXX}{Th\'{e}o Guyard, C\'{e}dric Herzet, and Cl\'{e}ment Elvira}

\ShortHeadings{\elops: An Exact L0-regularized Problems Solver}{Théo Guyard, Cédric Herzet, and Clément Elvira}
\firstpageno{1}

\title{\elops: An Exact L0-regularized Problems Solver}

\author{%
  \name Th\'{e}o Guyard \email theo.guyard@polymtl.ca\\
  \addr Scale-AI Chair, Polytechnique Montréal, Montréal, Canada\\
  \AND
  \name C\'{e}dric Herzet \email cedric.herzet@ensai.fr\\
  \addr Ensai, Université de Rennes, CREST - CNRS UMR 9194, Bruz, France\\
  \AND
  \name Cl\'{e}ment Elvira \email clement.elvira@centralesupelec.fr\\
  \addr CentraleSup\'{e}lec, Université de Rennes, IETR - CNRS UMR 6164, Cesson-S\'{e}vign\'{e}, France
}

\editor{}

\begin{document}

\maketitle

\begin{abstract}
  This paper presents \elops, a Python toolbox providing several utilities to handle $\ell_0$-regularized problems related to applications in machine learning, statistics, and signal processing, among other fields.
  In contrast to existing toolboxes, \elops allows users to define custom instances of these problems through a flexible framework, provides a dedicated solver achieving state-of-the-art performance, and offers several built-in machine learning pipelines.
  Our aim with \elops is to provide a comprehensive tool which opens new perspectives for the integration of $\ell_0$-regularized problems in practical applications.
\end{abstract}



\section{Introduction}

The past decade has seen a growing interest in optimization problems expressed as
\begin{equation}
  \label{prob:prob}
  \tag{$\ppb$}
  \textstyle
  \opt{\pobj} = \min_{\pv \in \kR^{\pdim}} \lfunc(\dic\pv) + \reg\norm{\pv}{0} + \sum_{\idxentry=1}^{\pdim}\pfunc(\pvi{\idxentry})
\end{equation}
seeking a minimization trade-off between a loss function $\lfunc$ composed with a matrix $\dic \in \kR^{\ddim\times\pdim}$, an $\ell_0$-norm regularization $\norm{\pv}{0}\triangleq\mathrm{card}(\kset{\idxentry}{\pvi{\idxentry} \neq 0})$ with weight $\reg > 0$ promoting sparse solutions, and penalty term $\pfunc$ applied coordinate-wisely to enforce other desirable properties.
These problems find applications in machine learning \citep{meng2024falcon}, statistics \citep{bertsimas2022solving}, and signal processing \citep{soussen2011bernoulli}, among many other fields \citep{tillmann2024cardinality}.
Despite recent methodological advances on their exact solution \citep{bertsimas2021unified,ben2022global,hazimeh2022sparse}, existing toolboxes remain limited in their numerical efficiency or their flexibility regarding the instances they can handle.

To address these limitations, we propose \elops, a Python package available at
\begin{center}
  \url{https://github.com/TheoGuyard/El0ps}
\end{center}
which offers three main improvements over existing toolboxes.
First, \elops is \emph{versatile} and allows user-defined instances of problem \eqref{prob:prob}, in addition to several common instances natively implemented.  
Second, \elops is \emph{efficient} and can address these instances via a dedicated \gls{bnb} solver achieving state-of-the-art performance.
Finally, \elops is \emph{ergonomic} and offers several convenient pipelines for practitioners.


\section{Hands-on with the Toolbox}
\label{sec:starting_with_elops_}

\paragraph{Instantiating Problems}

An instance of problem \eqref{prob:prob} is characterized by a tuple $(\lfunc{}, \pfunc{}, \dic, \reg{})$.
With \elops{}, users can define a matrix $\dic \in \kR^{\ddim \times \pdim}$ using any object compatible with \numpy{} \citep{harris2020array}, set $\reg > 0$ as a standard Python scalar, and choose from a variety of functions $\lfunc$ and $\pfunc$ natively implemented in the toolbox (see \Cref{tab:native-functions}).
A typical workflow example is provided in \Cref{fig:instantiate}.
Additionally, \elops{} allows for user-defined loss and penalty functions to better suit specific application needs.
Provided that they comply with the set of blanket assumptions specified in \Cref{sec:branch_and_bound_solver_implementation_choices}, they can be defined through convenient templates as detailed in \Cref{sec:instances}.
Besides the function itself, users only need to specify some basic operations such evaluating its conjugate, subdifferential, and proximal operator.

\begin{figure}[!ht]
    \centering
    \begin{minipage}{\codewidth\linewidth}
\begin{lstlisting}
import numpy as np
from el0ps.datafit import Logistic
from el0ps.penalty import L1L2norm

A = np.random.randn(100, 200)
y = np.random.choice([-1,1], size=100)
f = Logistic(y)
h = L1L2norm(alpha=0.5, beta=0.25)
lmbd = 0.1
\end{lstlisting}
    \end{minipage}
    \caption{Instantiation of components in problem \eqref{prob:prob} with a randomly-sampled matrix $\dic \in \kR^{100\times200}$, a logistic loss $\lfunc(\wv) = \sum_{j=1}^{100}\log(1 + \exp(-\wvi{j}\obsi{j}))$ for some randomly-sampled $\obs \in \{-1,+1\}^{100}$, an $\ell_1\ell_2^2$-norm penalty $\pfunc(\pvi{}) = \alpha\abs{\pvi{}} + \beta\pvi{}^2$ with $(\alpha,\beta)=(0.5,0.25)$, and an $\ell_0$-norm weight parameter $\reg = 0.1$.}
    \label{fig:instantiate}
\end{figure}

\paragraph{Solving Problems}

Once the different components of problem \eqref{prob:prob} have been instantiated, \elops offers a dedicated solver based on a generic \gls{bnb} framework introduced by \cite{Elvira:2025rf} whose main ingredients are outlined in \Cref{sec:bnb,sec:branch_and_bound_solver_implementation_choices}.
Different acceleration strategies proposed by \cite{hazimeh2022sparse}, \cite{samain2024techniques}, and \cite{theo2024new} are also implemented to enhance its performance.
As shown in \Cref{fig:solve}, the \gls{bnb} solver can be run with a single line of code.
It returns an output including the best solution and objective value found, as well as other information related to the solution process. 
The solver backbone is built upon \pybnb \citep{pybnb}, allowing for parallelization capabilities and a fine-tuning of different parameters governing the method behavior (stopping criterion, branching rule, exploration policies, ...).
It also broadly exploits \numba \citep{lam2015numba} to pre-compile code and accelerate matrix operations. 

\begin{figure}[!ht]
    \centering
    \begin{minipage}[h]{\codewidth\linewidth}
\begin{lstlisting}
from el0ps.solver import BnbSolver

solver = BnbSolver()
result = solver.solve(f, h, A, lmbd)
\end{lstlisting}
    \end{minipage}
    \caption{Solving problem \eqref{prob:prob} using the \gls{bnb} solver.}
    \label{fig:solve}
\end{figure}

\paragraph{Regularization Paths}

Beyond a mere solution method for problem \eqref{prob:prob}, \elops provides a pipeline to compute regularization paths \citep{friedman2010regularization}, that is, the solutions to problem \eqref{prob:prob} for different values of parameter $\reg$.
This generates a pool of candidates with varying sparsity levels --the larger $\reg$, the sparser the candidates-- among which practitioners can select one suited to their needs.
Users can directly specify the grid of values of $\reg$ to consider.
To assist the calibration of this range, \elops also includes a routine that automatically computes some quantity $\reg_{\max} > 0$ such that $\opt{\pv} = \0$ is a solution of any instance $(\lfunc, \pfunc, \dic, \reg)$ of problem~\eqref{prob:prob} with $\reg\geq\reg_{\max}$.
An example of regularization path construction is depicted in \Cref{fig:path}. 
The returned results are similar to those in \Cref{fig:solve}, but indexed by each value of $\reg$.

\begin{figure}[!ht]
    \centering
    \begin{minipage}[h]{\codewidth\linewidth}
\begin{lstlisting}
from el0ps.path import Path
from el0ps.utils import compute_lmbd_max

lmbd_max = compute_lmbd_max(f, h, A)
lmbd_min = 0.01 * lmbd_max
lmbd_num = 20

path = Path(lmbd_max, lmbd_min, lmbd_num)
results = Path.fit(solver, f, h, A)
\end{lstlisting}
    \end{minipage}
    \caption{Regularization path construction with $\reg_{\mathrm{num}} = 20$ values of $\reg$ spanned on a logarithmic grid over $[0.01 \times \reg_{\max},\reg_{\max}]$, where $\reg_{\max}$ is computed through a built-in routine.}
    \label{fig:path}
\end{figure}

\paragraph{Integration with \scikit}

To extend its usability, \elops comes with an interface to \scikit \citep{pedregosa2011scikit}.
By overloading the \texttt{LinearModel} class, it provides estimators of sparse linear models corresponding to solutions of $\ell_0$-regularized problems \citep{soussen2011bernoulli,tropp2010computational}, which are not natively implemented in \scikit.
This integration allows users to use $\ell_0$-regularization based estimators within the broad \scikit ecosystem, which includes cross-validation, hyperparameter tuning, and model evaluation pipelines, among others.
An example is given in \Cref{fig:scikit}.

\begin{figure}[!ht]
    \centering
    \begin{minipage}[h]{\codewidth\linewidth}
\begin{lstlisting}
from sklearn.model_selection import GridSearchCV
from el0ps.estimator import L0Estimator

estimator = L0Estimator(f, h, lmbd)

values = [0.01, 0.1, 1., 10.]
params = {'lmbd': values, 'alpha': values, 'beta': values}

grid_search_cv = GridSearchCV(estimator, params)
grid_search_cv.fit(A, y)
best_params = grid_search_cv.best_params_
\end{lstlisting}
    \end{minipage}
    \caption{Estimator corresponding to a solution of the same problem as instantiated in \Cref{fig:instantiate,fig:solve}, but used in a cross validation pipeline provided by \scikit to select the best combination of hyperparameters $(\reg, \regone, \regtwo)$, each tested among $\{0.01,0.1,1,10\}$.}
    \label{fig:scikit}
\end{figure}


\section{Comparison with Existing Toolboxes}
\label{sec:comparison_with_existing_toolboxes}

Prior to \elops, toolboxes able to solve problem \eqref{prob:prob} exactly were limited to instances where\footnote{$\icvx_{\mathcal{S}}$ denotes the indicator function of a set $\mathcal{S}$, defined as $\icvx_{\mathcal{S}}(\pvi{}) = 0$ if $\pvi{} \in \mathcal{S}$ and $\icvx_{\mathcal{S}}(\pvi{})=+\infty$ otherwise.} $\pfunc(\pvi{}) = \regtwo\pvi{}^2 + \icvx_{[-\bigM,\bigM]}(\pvi{})$ for some $\regtwo \in \kR+$ and $\bigM \in [0,+\infty]$, with at least $\regtwo > 0$ or $\bigM < +\infty$.
These instances can be cast into Mixed-Integer Programs \citep{hazimeh2022sparse} and tackled using generic solvers such as \gurobi \citep{gurobi}.\footnote{In this case, any term in the penalty not matching $\regtwo\pvi{}^2$ or $\icvx_{[-\bigM,\bigM]}(\pvi{})$ can be incorporated into the loss.}
An Outer Approximation method (\oa) was also proposed by \cite{bertsimas2021sparse} to enhance their performance and extend them to non-quadratic loss functions.
Besides, specialized \gls{bnb} solvers like \mimosa \citep{ben2022global} and \lobnb \citep{hazimeh2022sparse} specifically target least-squares losses, with \mimosa further requiring $\regtwo = 0$.
Finally, we mention that toolboxes can address \eqref{prob:prob} approximately via heuristics \citep{hazimeh2023l0learn} or relaxation strategies \citep{bertrand2022beyond}.

\begin{table}[!ht]
    \setlength{\tabcolsep}{4pt}
    \centering
    \scriptsize
    \begin{tabular}{cccc|ccccc}
        \toprule
        \textbf{Dataset} & \textbf{Dim. $\ddim \times \pdim$} & \textbf{Loss function} & \textbf{Penalty} & \gurobi & \oa & \lobnb & \mimosa & \elops \\
        \midrule 
        & & & $\regone = 0$, $\regtwo = 0$, $\bigM < +\infty$ & 1233.52 & 232.90 & 5.97 & 3.64 & \textbf{1.38} \\
        \texttt{Riboflavin} & $71 \times 4,088$ & Least-squares & $\regone = 0$, $\regtwo > 0$, $\bigM < +\infty$ & 1378.59 & 18.83 & 3.78 & \ding{54} & \textbf{0.79} \\
        & & & $\regone > 0$, $\regtwo > 0$, $\bigM < +\infty$ & 1027.32 & 2.31 & \ding{54} & \ding{54} & \textbf{0.61} \\
        \midrule
        & & & $\regone = 0$, $\regtwo = 0$, $\bigM < +\infty$ & \ding{54} & \rule[0.5ex]{0.5em}{2pt} & \ding{54} & \ding{54} & \textbf{1.31} \\
        \texttt{Leukemia} & $38 \times 7,129$ & Logistic & $\regone = 0$, $\regtwo > 0$, $\bigM < +\infty$ & \ding{54} & 13.52 & \ding{54} & \ding{54} &  \textbf{3.16} \\
        & & & $\regone > 0$, $\regtwo > 0$, $\bigM < +\infty$ & \ding{54} & 13.96 & \ding{54} & \ding{54} & \textbf{2.73} \\
        \midrule 
        & & & $\regone = 0$, $\regtwo = 0$, $\bigM < +\infty$ & 183.07 & \rule[0.5ex]{0.5em}{2pt} & \ding{54} & \ding{54} & \textbf{10.37} \\
        \texttt{Arcene} & $100 \times 10,000$ & Squared-hinge & $\regone = 0$, $\regtwo > 0$, $\bigM < +\infty$ & \rule[0.5ex]{0.5em}{2pt} & \rule[0.5ex]{0.5em}{2pt} & \ding{54} & \ding{54} & \textbf{27.76} \\
        & & & $\regone > 0$, $\regtwo > 0$, $\bigM < +\infty$ & 256.46 & 47.59 & \ding{54} & \ding{54} & \textbf{11.44} \\
        \bottomrule
    \end{tabular}
    \caption{Solving time (sec.) for feature selection tasks. ``\ding{54}'' indicates solvers that could not handle  instances of problem \eqref{prob:prob}. ``\rule[0.5ex]{0.5em}{2pt}'' indicates solvers that did not terminate within 1 hour.}
    \label{tab:numerics}
\end{table}

In \Cref{tab:numerics}, we consider machine learning datasets from OpenML \citep{OpenML2013}, each associated with a feature matrix $\dic \in \kR^{\ddim\times\pdim}$ and target vector $\obs \in \kR^{\ddim}$.
We perform feature selection tasks by solving instances of \eqref{prob:prob} with an appropriate loss function and three variations of the penalty $\pfunc(\pvi{}) = \regone\abs{\pvi{}} + \regtwo\pvi{}^2 + \icvx_{[-\bigM,\bigM]}(\pvi{})$, which benefits from desirable statistical properties \citep{dedieu2021learning}. 
The quantities $(\reg,\regone,\regtwo,\bigM)$ are calibrated using the grid search procedure detailed in \Cref{sec:supp_numerics}.
Our results outline that \elops outperforms its competitors both in terms of variety in instances it can handle and solving time.


\section{Conclusion}
\label{sec:conclusion}

This paper introduces \elops{}, a Python toolbox providing several utilities for $\ell_0$-regularized problems.
It stands out from existing toolboxes thanks to its flexible instantiation framework, its state-of-the-art solver, and the pipelines it offers for practitioners.
Our belief with \elops{} is to open new perspectives for integrating $\ell_0$-regularized problems in real-world applications.

\acks{Part of this work has been carried out when Th\'{e}o Guyard was affiliated to INSA Rennes, IRMAR - CNRS UMR 6625, France and funded by the ANR-11-LABX-0020.}
\appendix
\bibliography{bibliography}
\clearpage

\paragraph{Notations}
In the following appendices, we denote by $\kR$ the set of real numbers and by $\kR^{\pdim}$ the set of $\pdim$-dimensional real vectors. 
Scalars are denoted by lowercase letters (\eg, $x$), vectors by bold lowercase letters (\eg, $\mathbf{x}$), and matrices by bold uppercase letters (\eg, $\mathbf{X}$).
For any vector $\pv\in\kR^\pdim$, we note $\pvi{\idxentry{}}$ its $\idxentry{}$-th entry and $\pv{}_{\nodesymbol{}}$ its restriction to its entries indexed by $\nodesymbol{}\subseteq \intervint{1}{\pdim}$.
Similarly, for any matrix $\mathbf{X}\in\kR^{\pdim\times\pdim}$, we denote by $\mathbf{x}_{\idxentry}$ the restriction to its $\idxentry$-th column and by $\mathbf{X}_{\nodesymbol{}}$ the restriction to its columns indexed by $\nodesymbol{}$.
Given some set $\mathcal{X}\subseteq \kR^\pdim$, $\interior(\mathcal{X})$ refers to its interior \cite[Sec.~1.7]{bauschke2017convex} and its indicator $\icvx_{\mathcal{X}}$ is defined as $\icvx_{\mathcal{X}}(\pv) = 0$ if $\pv \in \mathcal{X}$ and $\icvx_{\mathcal{X}}(\pv) = +\infty$ otherwise.
Moreover, the positive part function is defined as $\pospart{\pvi{}} = \max(\pvi{},0)$ for all $\pvi{} \in \kR$.
Finally, given some function $\kfuncdef{f}{\mathcal{X}}{[-\infty,+\infty]}$, we let $\dom f$ its domain, and the notations $\prox_{f}$, $\conj{f}$, $\biconj{f}$, and $\subdiff{f}$ respectively refer to the proximal operator, convex conjugate, convex bi-conjugate, and subdifferential associated with $f$ \cite[Defs.~12.23, 13.1, 16.1]{bauschke2017convex}. 


\section{Branch-and-Bound Solver: Main Principles}
\label{sec:bnb} 

In this section, we present the main principles of a specialized \glsreset{bnb}\gls{bnb} procedure tailored to problem \eqref{prob:prob}.
The specific choices made in \elops for the implementation of this method are further detailed in \Cref{sec:branch_and_bound_solver_implementation_choices}.

\subsection{Algorithmic Principles}
\label{sec:bnb:principle}

A \gls{bnb} algorithm addresses an optimization problem by exploring regions of its feasible space and pruning those that cannot contain optimal solutions \citep{lawler1966branch}.
This general principle can be specialized to problem \eqref{prob:prob}.
More precisely, a region in its feasible space $\kR^{\pdim}$ can be identified by two disjoint subsets \(\nodesymbol = (\setzero,\setone)\) of \(\intervint{1}{\pdim}\) and defined as
\begin{align}
    \label{eq:region}
    \node{\mathcal{X}}
    &= \kset{\pv\in\kR^\pdim}{\subzero{\pv} = \0, \ \subone{\pv} \neq \0}
    .
\end{align}
If some \textit{``pruning test''} is passed (see \Cref{sec:bnb:pruning}) for a given region, the latter does not contain any optimal solution and can be discarded from the optimization problem.
Otherwise, the region is further partitioned into two new regions (see \Cref{sec:bnb:exploration}). 
When the sets \(\nodesymbol = (\setzero,\setone)\) partition \(\intervint{1}{\pdim}\), a local solution to problem \eqref{prob:prob} over the corresponding region $\node{\mathcal{X}}$ can be computed with a tractable complexity since the $\ell_0$-norm in problem \eqref{prob:prob} vanishes.
The process is repeated until the whole feasible space has been explored, after which the optimal solutions are identified among the local ones obtained during the procedure.

\subsection{Pruning Tests}
\label{sec:bnb:pruning}

A pruning test is performed for each region $\node{\mathcal{X}}$ explored during the \gls{bnb} algorithm. 
It aims to determine whether this region can contain some optimal solution to~\eqref{prob:prob}. 
Recalling that $\opt{\pobj}$ denote the optimal value of problem \eqref{prob:prob} and letting
\begin{equation}
    \label{prob:node-prob}
    \tag{$\node{\ppb}$}
    \textstyle
    \node{\pobj} = 
    \inf_{\pv \in \node{\mathcal{X}}} \lfunc(\dic\pv) + \reg \|\pv{}\|_0 
    + \sum_{\idxentry{}=1}^{\pdim}\pfunc(\pvi{\idxentry{}})
    ,
\end{equation}
the pruning test amounts to checking whether $\node{\pobj} > \opt{\pobj}$.
Since the computation of these quantities is intractable, one rather implements a pruning test involving upper and lower bounds as surrogates.
Specifically, if $\node{\lbpobj} \leq \node{\pobj}$ and $\UB{\pobj} \geq \opt{\pobj}$, we have 
\begin{align}
    \label{eq:pruning-test}
    \node{\lbpobj} > \UB{\pobj} \implies \node{\pobj} > \opt{\pobj}
    \implies \node{\mathcal{X}} \cap \opt{\mathcal{X}} = \emptyset
    ,
\end{align}
where $\opt{\mathcal{X}}$ denotes the set of optimal solutions to problem \eqref{prob:prob}.
\Cref{sec:elops:pruning_test} specifies the implementation choices made in \elops to compute these bounds.

Besides standard pruning tests which only focus on the region considered at the current stage of the \gls{bnb} algorithm, \cite{theo2024new} introduced a \textit{``simultaneous''} pruning strategy directly tailored to problem \eqref{prob:prob}.
More precisely, let $\nodesymbol = (\setzero,\setone)$ and denote
\begin{align}
    \label{eq:childnode0}
    \childnode{\nodesymbol}{\idxentry}{0} 
    &= 
    (\setzero\cup\{\idxentry{}\}, \setone)\\
    \label{eq:childnode1}
    \childnode{\nodesymbol}{\idxentry}{1} 
    &= 
    (\setzero, \setone\cup\{\idxentry{}\})
\end{align}
for any $\idxentry{} \in \{1,\ldots,\pdim{}\}\setminus(\setzero{}\cup\setone{})$.
While assessing region $\node{\mathcal{X}}$ using standard pruning operations, the simultaneous pruning method allows testing (at virtually no cost) any region $\mathcal{X}^{\nodesymbol{}'}$ where
\begin{align}
    \label{eq:bnb:set-nodes-simul-pruning}
    \nodesymbol{}' \in \kset{\childnode{\nodesymbol}{\idxentry}{b}}{\idxentry{} \in \{1,\ldots,\pdim{}\}\setminus(\setzero{}\cup\setone{}), b \in \{0,1\}}
    .
\end{align} 
\elops implements this simultaneous pruning strategy following the principles described in \cite[Sec.~3.4]{theo2024new}.

\subsection{Branching and Exploration Strategies}
\label{sec:bnb:exploration}

At each step of the \gls{bnb} algorithm, the procedure must decide which region $\node{\mathcal{X}}$ to explore—that is, to test for pruning and, if the test fails, to partition further. The method used to select the next region to explore is commonly referred to as the \textit{``exploration strategy''}.

If a region $\node{\mathcal{X}}$ passes no pruning test, the \gls{bnb} procedure partitions it into two new regions,
$\mathcal{X}^{\childnode{\nodesymbol}{\idxentry}{0}}$ and $\mathcal{X}^{\childnode{\nodesymbol}{\idxentry}{1}}$, where the pair $(\childnode{\nodesymbol}{\idxentry}{0}, \childnode{\nodesymbol}{\idxentry}{1})$ is defined as in \eqref{eq:childnode0}–\eqref{eq:childnode1} for some $\idxentry{} \in \{1,\ldots,\pdim{}\} \setminus (\setzero{} \cup \setone{})$.
The choice of the index $\idxentry{}$ is typically referred to as the \textit{``branching strategy''}.

The exploration and branching strategies implemented in \elops{} are described in \Cref{sec:elops:exploration}. 


\section{Branch-and-Bound Solver: Implementation Choices}
\label{sec:branch_and_bound_solver_implementation_choices}

In this section, we describe the choices made in \elops for the implementation of the \glsreset{bnb}\gls{bnb} procedure described in \Cref{sec:bnb}.

\subsection{Working Assumptions and Key Quantities}

\elops is designed to handle any instance of problem \eqref{prob:prob} verifying the following assumptions:
\begin{enumerate}[label=\textnormal{({H\arabic*})},itemsep=4pt,topsep=12pt,parsep=0pt,leftmargin=40pt, start=0]
    \item \label{assumption:f}
        \textit{\(\lfunc\) closed, convex, differentiable, lower-bounded, and $\0 \in \interior(\dom\lfunc)$}.
    \item \label{assumption:zero-minimized}
        \textit{\(\pfunc(\pvi{}) \geq \pfunc(0)=0\) for all $\pvi{} \in \kR$ and \(\dom\pfunc\cap\kR+\setminus\{0\}\neq\emptyset\)}.
     \item \label{assumption:closed}    
        \textit{\(\pfunc\) is closed}. 
    \item \label{assumption:convex}
        \textit{\(\pfunc\) is convex.}              
    \item \label{assumption:coercive}
        \textit{\(\pfunc\) is coercive.}
    \item \label{assumption:even}
        \textit{\(\pfunc\) is even.} 
\end{enumerate}
In this framework, \cite{Elvira:2025rf} established useful results for the construction of the \gls{bnb} algorithm presented in \Cref{sec:bnb}.\footnote{\elops also accepts penalties that do not satisfy assumption \ref{assumption:even} via a straightforward extension of the work of \cite{Elvira:2025rf}. However, we still consider this assumption to lighten our exposition.}
In particular, \cite[Proposition 1]{Elvira:2025rf} shows that problem \eqref{prob:node-prob} can be reformulated as 
\begin{equation}
    \label{prob:prob-node}
    \textstyle
    \node{\pobj} = \min_{\pv \in \kR^{\pdim}} \lfunc(\dic\pv) 
    + 
    \sum_{\idxentry=1}^{\pdim} \node{\separable{\rfunc}{\idxentry}}(\pvi{\idxentry})
\end{equation}
with
\begin{equation}
    \label{eq:regfunc-node} 
    \node{\separable{\rfunc}{\idxentry}}(\pvi{}) = 
    \begin{cases}
        \icvx_{\{0\}}(\pvi{}) &\text{if} \ \idxentry \in \setzero \\
        \pfunc(\pvi{}) + \reg &\text{if} \ \idxentry \in \setone \\
        \rfunc(\pvi{}) &\text{otherwise}
    \end{cases}
\end{equation}
and where $\rfunc(\pvi{}) = \pfunc(\pvi{}) + \reg \|\pvi{}\|_0$.
As emphasized in \cite[Sections 3.2 to 3.4]{Elvira:2025rf}, all the quantities of interest for the \gls{bnb} implementation are related to the convex conjugate and bi-conjugate functions of $\rfunc$.
For instance, it is established that the bounds involved in the pruning test \eqref{eq:pruning-test} can be derived from convex optimization problems involving the tightest convex relaxation of $\node{\separable{\rfunc}{\idxentry}}$ and its conjugate, respectively given by
\begin{equation}
    \label{eq:relax-gi}
    \node{\separable{\relaxregfunc}{\idxentry}}(\pvi{}) =
    \begin{cases}
        \icvx_{\{0\}}(\pvi{}) &\text{if} \ \idxentry \in \setzero \\
        \pfunc(\pvi{}) + \reg &\text{if} \ \idxentry \in \setone \\
        \biconj{\rfunc}(\pvi{}) &\text{otherwise,}
    \end{cases}         
    \qquad\text{and}\qquad
    \conj{(\node{\separable{\relaxregfunc}{\idxentry}})}(\bvi{}) = 
    \left\{
        \begin{array}{ll}
            0 & \text{if} \ \idxentry \in \setzero \\
            \conj{\pfunc}(\bvi{}) - \reg & \text{if} \ \idxentry \in \setone 
            \\
            \conj{\rfunc}(\bvi{}) &\text{otherwise.} 
        \end{array}
    \right.     
\end{equation} 
Finally, it is shown in \cite[Section 4]{Elvira:2025rf} that closed-form expressions for the functions $\conj{\rfunc}$ and $\biconj{\rfunc}$ appearing in \eqref{eq:relax-gi}, as well as their subdifferential and proximal operators, admits closed-form expressions only depending on the following quantities:
\begin{align}
    \label{eq:pertslope}
    \pertslope &=  \sup \kset{\bvi{}\in\kR+}{\conj{\pfunc}(\bvi{}) \leq \reg} 
    \\
    \label{eq:pertlimit}
    \pertlimit &= 
    \begin{cases}
        \sup \kset{\bvi{}\in\kR+}{\bvi{} \in \subdiff\conj{\pfunc}(\pertslope)} &\text{ if } \subdiff\conj{\pfunc}(\pertslope) \neq \emptyset \\
        +\infty &\text{ otherwise}
    \end{cases}
    \\
    \label{eq:pertrightlope}
    \pertrightlope &=
    \begin{cases}
     \sup\kset{\bvi{}\in\kR+}{\bvi{} \in \subdiff\pfunc(\pertlimit)} & \mbox{if \(\pertlimit<+\infty\)}\\
     +\infty &\mbox{otherwise.}
     \end{cases}
\end{align}
As further explained in \Cref{sec:instances:user-defined}, users can implement their own penalty functions $\pfunc$. 
In this case, a closed-form expression for the parameter $\pertslope$ can be provided.
Otherwise, \elops{} automatically approximates it using a bisection method up to some prescribed precision.
By default, the parameters $(\pertlimit,\pertrightlope)$ are derived numerically from $\subdiff\pfunc$ and $\subdiff\conj{\pfunc}$, but can also be provided in closed-form.

\subsection{Pruning Tests}
\label{sec:elops:pruning_test}

In the following, we describe the implementation choices made in \elops for the computation of the pruning tests discussed in \Cref{sec:bnb:pruning}.  

\subsubsection{Upper Bound}

\elops{} updates the upper bound appearing in pruning test \eqref{eq:pruning-test} each time a new region $\node{\mathcal{X}}$ is explored during the \gls{bnb} procedure. The update reads $\UB{\pobj} \leftarrow \min(\UB{\pobj},\node{\UB{\pobj}})$ where
\begin{equation}
  \label{prob:node-heur}
  \tag{$\node{\ppb}_{\text{ub}}$}
  \textstyle
  \node{\UB{\pobj}} = \min_{\pv \in \kR^{\pdim}} \lfunc(\dic\pv) 
  + 
  \sum_{\idxentry=1}^{\pdim} \node{\separable{\relaxregfunc}{\idxentry}}(\pvi{}) 
  \quad\text{s.t.}\quad
    \pvi{\idxentry} = 0 \ \ \forall \idxentry \notin \setone{}
\end{equation}
with $\node{\separable{\relaxregfunc}{\idxentry}}$ defined in \eqref{eq:relax-gi}.
Under assumptions \ref{assumption:f} and \ref{assumption:convex}, the optimization problem \eqref{prob:node-heur} is convex.
The procedure implemented in \elops{} to tackle it is detailed in \Cref{sec:elops:bounding_solver}.
It is only run to some prescribed accuracy to avoid a too large computation load during the \gls{bnb} procedure.
However, we note that this does not alter the validity of the pruning process since the value of cost function in \eqref{prob:node-heur} at any point provides a valid upper bound on $\opt{\pobj}$.

\subsubsection{Lower Bound}
\label{sec:elops:lower_bound}

The lower bound $\node{\lbpobj}$ associated with some $\nodesymbol{}=(\setzero{},\setone{})$ is computed by \elops{} as
\begin{align}
    \label{eq:elops:cvx-node-prob}
    \tag{$\node{\ppb}_{\text{lb}}$}
    \textstyle
    \node{\lbpobj} = \min_{\pv \in \kR^{\pdim}} 
    \lfunc(\dic\pv) 
    + 
    \sum_{\idxentry=1}^{\pdim} \node{\separable{\relaxregfunc}{\idxentry}}(\pvi{})
\end{align}
where $\node{\separable{\relaxregfunc}{\idxentry}}$ is defined in \eqref{eq:relax-gi}.
Under assumptions \ref{assumption:f} and \ref{assumption:convex}, problem \eqref{eq:elops:cvx-node-prob} is convex.
The procedure implemented in \elops{} to address it is detailed in \Cref{sec:elops:bounding_solver}.
As for the upper bounding problem, \elops{} only solves problem \eqref{eq:elops:cvx-node-prob} to some prescribed accuracy to avoid a too large computation load during the \gls{bnb} procedure.
For this lower-bound problem however, this may alter the validity of the pruning process.
To preserve the correctness of the \gls{bnb} algorithm, we adopt the strategy described in \cite[Sec.~3.3]{Elvira:2025rf}.
For each iterate $\tilde{\pv} \in \kR^{\pdim}$ generated by the numerical procedure solving \eqref{eq:elops:cvx-node-prob}, \elops evaluates
\begin{align}
    \tilde{\dv} = -\grad\lfunc(\dic\tilde{\pv}) 
\end{align}
and computes the quantity
\begin{equation}
    \label{eq:elops:lbound}
    \textstyle
    \node{\dfunc}(\tilde{\dv}) =-\conj{\lfunc}(-\tilde{\dv}) - \sum_{\idxentry=1}^{\pdim} \conj{(\node{\separable{\relaxregfunc}{\idxentry}})}(\ktranspose{\atom{}}_\idxentry \tilde{\dv})
\end{equation}
where $\conj{(\node{\separable{\relaxregfunc}{\idxentry}})}$ is defined by \eqref{eq:relax-gi}.
Due to the weak Fenchel-Rockafellar duality property \cite[Chap.~19]{bauschke2017convex}, we have 
\begin{equation}
    \node{\dfunc}(\tilde{\dv}) \leq \node{\lbpobj} \leq \node{\pobj}
\end{equation}
which ensures that $\node{\dfunc}(\tilde{\dv})$ can always be used as surrogate for $\node{\lbpobj}$ in the pruning test \eqref{eq:pruning-test} when problem \eqref{eq:elops:cvx-node-prob} is not solved to machine accuracy.
This quantity is also involved in the implementation of the simultaneous pruning evaluated in \elops, implemented according to the approach proposed by \cite{theo2024new}.

\subsection{Exploration Strategy and Branching}
\label{sec:elops:exploration}

When the numerical procedure solving the lower-bounding problem \eqref{eq:elops:cvx-node-prob} runs until completion without any pruning test being successful, the \gls{bnb} algorithm must partition the current region and select a new one to explore, as described in \Cref{sec:bnb:exploration}.
On the one hand, \elops{} implements a branching strategy inspired from that proposed by \cite{mhenni2020sparse}.
Specifically, for a region $\node{\mathcal{X}}$ associated with some $\nodesymbol=(\setzero{},\setone{})$, and index
\begin{equation}
    \idxentry \in \argmax_{j \notin \setzero\cup\setone} \abs{\node{\pvi{j}}}
\end{equation}
is selected based on the solution $\node{\pv} \in \kR^{\pdim}$ obtained from problem \eqref{eq:elops:cvx-node-prob}.
Then, the region is partitioned into two sub-regions $\mathcal{X}^{\childnode{\nodesymbol}{\idxentry}{0}}$ and $\mathcal{X}^{\childnode{\nodesymbol}{\idxentry}{1}}$ as defined in \eqref{eq:childnode0}-\eqref{eq:childnode1}.
On the other hand, the default exploration strategy implemented in \elops selects the next region to explore is based on the \textit{``Best-First Search''} paradigm, which prioritizes regions with the smallest lower bound in their associated pruning tests.
Several other standard exploration strategies provided by the \pybnb framework upon which \elops is built are also available. We refer the reader to \pybnb's documentation for more details.

\subsection{Numerical Solver for Bounding Problems}
\label{sec:elops:bounding_solver}

\elops addresses both bounding problems \eqref{prob:node-heur} and \eqref{eq:elops:cvx-node-prob} via an efficient convex optimization methods.
We note from \eqref{eq:relax-gi} that \eqref{prob:node-heur} corresponds to a particular instance of \eqref{eq:elops:cvx-node-prob}.
Hence, we only detail the solving process of the lower bounding problem to simplify our exposition.

\subsubsection{Working Set Algorithm}
Inspired by the broad literature on sparse convex optimization \citep{johnson2015blitz,hazimeh2022sparse,bertrand2022beyond}, \elops implements a working set method to address problem \eqref{eq:elops:cvx-node-prob}.
Starting with an initial working set $\workingset \subseteq \{1,\dots,\pdim\}$, the procedure repeats the following steps until a stopping criterion is met.
\begin{itemize}
    \item \textbf{Step 1:} Compute a solution $\tilde{\pv} \in \kR^{\pdim}$ up to some prescribed accuracy of the sub-problem
    \begin{equation}
        \label{prob:convex-subproblem}
        \textstyle
        \min_{\pv \in \kR^{\pdim}} \lfunc(\dic\pv) + \sum_{\idxentry=1}^{\pdim} \node{\separable{\relaxregfunc}{\idxentry}}(\pvi{})
        \quad\text{s.t.}\quad
        \pvi{\idxentry} = 0, \ \forall \idxentry \notin \workingset
    \end{equation}
    where entries that do not belong to the current working set $\workingset$ are fixed to zero.
    \item \textbf{Step 2:} Compute $\tilde{\dv} = -\grad\lfunc(\dic\tilde{\pv})$ and use it to evaluate a pruning test based on the quantity $\node{\dfunc}(\tilde{\dv})$ as described in \Cref{sec:elops:lower_bound}. If it is verified, the procedure is stopped and the current region $\node{\mathcal{X}}$ is directly pruned.
    \item \textbf{Step 3:} Compute the set of indices
    \begin{equation}
        \label{eq:elops:violation-set}
        \violationset = \kset{\idxentry \notin \workingset}{0 \notin \transp{\dici{\idxentry}}\tilde{\dv} + \subdiff\node{\separable{\relaxregfunc}{\idxentry}}(\tilde{\pvi{}}_{\idxentry})}
    \end{equation}
    which derives from Fermat's optimality condition \cite[Thm.~16.3]{bauschke2017convex}. When $\violationset = \emptyset$, then $\tilde{\pv} \in \kR^{\pdim}$ corresponds to an optimal solution of problem \eqref{eq:elops:cvx-node-prob} and the procedure terminates. Otherwise, the working set is expanded as $\workingset \leftarrow \workingset \cup \violationset$ and the algorithm returns to Step 1.
\end{itemize}
This working set method allows saving computations since only a part of the optimization variable is considered when solving each sub-problem \eqref{prob:convex-subproblem}.
In \elops, the initial working set used for the regions $\mathcal{X}^{\childnode{\nodesymbol}{\idxentry}{0}}$ and $\mathcal{X}^{\childnode{\nodesymbol}{\idxentry}{1}}$ constructed as discussed in \Cref{sec:bnb:principle} is warm-started from the last one used in the region $\mathcal{X}^{\nodesymbol}$ from which these subregions originate.
Moreover, Step 1 is performed using a Coordinate Descent method \citep{wright2015coordinate}.
All quantities required for its implementation are automatically derived by \elops based on the work of \cite{Elvira:2025rf}.

\subsubsection{Screening Tests}
\label{sec:inner:screening}

To improve the working set selection, \elops{} never checks entries in $\setzero$ for the update \eqref{eq:elops:violation-set} since the latter must be zero at optimality according to the definition of $\subdiff\node{\separable{\relaxregfunc}{\idxentry}}$ given in \eqref{eq:relax-gi}, and therefore need not be included in the working set.
Whenever $\lfunc$ has an $\lipschitz$-Lipschitz continuous gradient, \elops also implements screening tests \citep{ndiaye2021screening} to identify other entries that need not be included in the working set.
More precisely, let 
\begin{equation}
        \spherecenter = \tfrac{1}{2}(\dv - \grad\lfunc(\dic\pv)) \quad\text{and}\quad
        \sphereradius = \sqrt{\lipschitz\gap(\pv,\dv) - \tfrac{1}{4}\norm{\dv + \grad\lfunc(\dic\pv)}{2}^2}
\end{equation}
where $\lipschitz$ is the Lipschitz-constant of $\grad\lfunc$ and $\gap(\pv,\dv)$ denotes the difference between the objective of problem \eqref{eq:elops:cvx-node-prob} evaluated at some $\pv \in \kR^{\pdim}$ and the quantity $\node{\dfunc}(\dv)$ defined in \eqref{eq:elops:lbound} for some $\dv \in \kR^{\ddim}$. 
Then, applying the screening test strategy proposed by \cite{tran2023one} to our setting yields that
\begin{subequations}
    \begin{alignat}{4}
        &\interior([\transp{\dici{\idxentry}}\spherecenter - \sphereradius, \transp{\dici{\idxentry}}\spherecenter + \sphereradius]) \subseteq \subdiff\pfunc(0) &&\implies \node{\pvi{\idxentry}} = 0 &&\quad\text{for all } \idxentry \in \setone
        \\
        &\interior([\transp{\dici{\idxentry}}\spherecenter - \sphereradius, \transp{\dici{\idxentry}}\spherecenter + \sphereradius]) \subseteq \subdiff\biconj{\rfunc}(0) &&\implies \node{\pvi{\idxentry}} = 0 &&\quad\text{for all } \idxentry \in \intervint{1}{\pdim} \setminus (\setzero \cup \setone)
    \end{alignat}
\end{subequations}
for any solution $\node{\pv} \in \kR^{\pdim}$ to problem \eqref{eq:elops:cvx-node-prob}.
\elops assesses these implications after each iteration of Step 2 using $(\pv,\dv)=(\tilde{\pv},\tilde{\dv})$ and filters entries when computing \eqref{eq:elops:violation-set} by removing those that must be zero at optimality since they need not be included in the working set.


\section{Loss and Penalty Functions}
\label{sec:instances}

This section gives additional details on the loss and penalty functions available in \elops, with two companion examples provided in \Cref{fig:custom-loss,fig:custom-penalty}.

\subsection{Native Functions}

\elops natively includes several loss and penalty functions commonly found in applications.
\Cref{tab:native-functions} summarizes the different classes corresponding to these functions and specifies the parameters involved in their instantiation.

\begin{table}[!ht]
    \centering
    \small
    \begin{tabular}{llll}
        \toprule
        \textbf{Class} & \textbf{Parameters} & \textbf{Expression} \\
        \midrule
        \texttt{KullbackLeibler} & \texttt{y:NDArray, eps:float} & $\lfunc(\wv) = \sum_{j=1}^{\ddim} \obsi{j}\log(\tfrac{\obsi{j}}{\wvi{j} + \epsilon}) + \wvi{j} + \epsilon - \obsi{j}$  \\
        \texttt{Leastsquares} & \texttt{y:NDArray} & $\lfunc(\wv) = \sum_{j=1}^{\ddim} \tfrac{1}{2}(\wvi{j} - \obsi{j})^2$ \\
        \texttt{Logcosh} & \texttt{y:NDArray} & $\lfunc(\wv) = \sum_{j=1}^{\ddim} \log(\cosh(\wvi{j} - \obsi{j}))$ \\
        \texttt{Logistic} & \texttt{y:NDArray} & $\lfunc(\wv) = \sum_{j=1}^{\ddim} \log(1 + \exp(- \wvi{j}\obsi{j}))$ \\
        \texttt{Squaredhinge} & \texttt{y:NDArray} & $\lfunc(\wv) = \sum_{j=1}^{\ddim} \pospart{1 - \wvi{j}\obsi{j}}^2$ \\
        \midrule
        \texttt{Bigm} & \texttt{M:float} & $\pfunc(\pvi{}) = \icvx_{[-\bigM,\bigM]}(\pvi{})$ \\
        \texttt{BigmL1norm} & \texttt{M:float, alpha:float} & $\pfunc(\pvi{}) = \icvx_{[-\bigM,\bigM]}(\pvi{}) + \alpha\abs{\pvi{}}$ \\
        \texttt{BigmL1norm} & \texttt{M:float, beta:float} & $\pfunc(\pvi{}) = \icvx_{[-\bigM,\bigM]}(\pvi{}) + \beta\pvi{}^2$ \\
        \texttt{BigmPositiveL1norm} & \texttt{M:float, alpha:float} & $\pfunc(\pvi{}) = \icvx_{[0,\bigM]}(\pvi{}) + \alpha\pvi{}$ \\
        \texttt{BigmPositiveL2norm} & \texttt{M:float, beta:float} & $\pfunc(\pvi{}) = \icvx_{[0,\bigM]}(\pvi{}) + \beta\pvi{}^2$ \\
        \texttt{Bounds} & \texttt{x\_lb:float, x\_ub:float} & $\pfunc(\pvi{}) = \icvx_{[\pvi{\text{lb}},\pvi{\text{ub}}]}(\pvi{})$ \\
        \texttt{L1L2norm} & \texttt{alpha:float, beta:float} & $\pfunc(\pvi{}) = \alpha\abs{\pvi{}} + \beta\pvi{}^2$ \\
        \texttt{L1norm} & \texttt{alpha:float} & $\pfunc(\pvi{}) = \alpha\abs{\pvi{}}$ \\
        \texttt{L2norm} & \texttt{beta:float} & $\pfunc(\pvi{}) = \beta\pvi{}^2$ \\
        \texttt{PositiveL1norm} & \texttt{alpha:float} & $\pfunc(\pvi{}) = \icvx_{[0,+\infty)}(\pvi{}) + \alpha\abs{\pvi{}}$ \\
        \texttt{PositiveL2norm} & \texttt{beta:float} & $\pfunc(\pvi{}) = \icvx_{[0,+\infty)}(\pvi{}) + \beta\pvi{}^2$ \\
        \bottomrule
    \end{tabular}
    \caption{Loss and penalty functions natively implemented in \elops. Parameter \texttt{y} is required to be a one-dimensional \numpy-compatible array (\texttt{NDArray}). All \texttt{float} parameters are required to be positive, except \texttt{x\_lb} which is required to be negative. In the \texttt{KullbackLeibler} loss, the logarithm function is extended as $\log(\wvi{}) = +\infty$ whenever $\wvi{} \leq 0$. }
    \label{tab:native-functions}
\end{table}

\subsection{User-defined Functions}
\label{sec:instances:user-defined}

A key feature of \elops is to allow user-defined loss and penalty functions complying with the set of blanket assumptions specified in \Cref{sec:branch_and_bound_solver_implementation_choices}, in addition those natively implemented by the package.
This is done by overloading some template classes.

\paragraph{User-defined Loss}
A custom loss $\lfunc$ can be implemented by deriving from the \texttt{BaseDatafit} class, which requires specifying the following methods:
\begin{itemize}
    \item \texttt{value(self, w:NDArray)}: value of the function $\lfunc$ at $\wv \in \kR^{\ddim}$.
    \item \texttt{conjugate(self, w:NDArray)}: value of the conjugate function $\conj{\lfunc}$ at $\wv \in \kR^{\ddim}$.
    \item \texttt{gradient(self, w:NDArray)}: value of the gradient $\grad\lfunc$ at $\wv \in \kR^{\ddim}$.
    \item \texttt{gradient\_lipschitz\_constant(self)}: Lipschitz-constant of the gradient $\grad\lfunc$, if any.
\end{itemize}
A constructor can also be defined to specify parameters for the loss at instantiation.

\paragraph{User-defined Penalty}
A custom penalty $\pfunc$ can be implemented by deriving from the \texttt{SymmetricPenalty} class, which requires specifying the following methods:
\begin{itemize}
    \item \texttt{value(self, i:int, x:float)}: value of the function $\pfunc$ at $\pvi{\idxentry} \in \kR$.
    \item \texttt{conjugate(self, i:int, x:float)}: value of the conjugate function $\conj{\pfunc}$ at $\pvi{\idxentry} \in \kR$.
    \item \texttt{prox(self, i:int, x:float, eta:float)}: value of $\prox_{\eta\lfunc}$ at $\pvi{\idxentry} \in \kR$ for some $\eta > 0$.
    \item \texttt{subdiff(self, i:int, x:float)}: bounds defining the set $\subdiff{\pfunc}$ at $\pvi{\idxentry} \in \kR$.
    \item \texttt{conjugate\_subdiff(self, i:int, x:float)}: bounds defining the set $\subdiff\conj{\pfunc}$ at $\pvi{\idxentry} \in \kR$.
\end{itemize}
By default, \elops approximates numerically the value of the parameters $(\pertslope,\pertlimit,\pertrightlope)$ defined in \eqref{eq:pertslope}-\eqref{eq:pertlimit}-\eqref{eq:pertrightlope} and involved in \gls{bnb} operations for user-defined penalty functions.
However, they can also be provided in closed form through the following methods:
\begin{itemize}
    \item \texttt{param\_slope(self, i:int, lmbd:float)}: value of $\pertslope$ for some $\reg > 0$.
    \item \texttt{param\_limit(self, i:int, lmbd:float)}: value of $\pertlimit$ for some $\reg > 0$.
    \item \texttt{param\_bndry(self, i:int, lmbd:float)}: value of $\pertrightlope$ for some $\reg > 0$.
\end{itemize}
A constructor method can also be defined to specify parameters of the penalty at instantiation.
We note that users can easily define a function $\pfunc$ that takes a different expression depending on the index $\idxentry \in \{1,\dots,\pdim\}$ of the variable $\pvi{\idxentry}$ considered based on the argument \texttt{i:int} in the above methods.
This corresponds to trading $\sum_{\idxentry=1}^{\pdim}\pfunc(\pvi{\idxentry})$ for $\sum_{\idxentry=1}^{\pdim}\separable{\pfunc}{\idxentry}(\pvi{\idxentry})$ in problem \eqref{prob:prob}, where splitting terms $\{\separable{\pfunc}{\idxentry}\}_{\idxentry=1}^{\pdim}$ can have different expressions.

\paragraph{Non-symmetric Penalty}
User-defined penalty functions that do not verify assumption \ref{assumption:even} are also supported by \elops.
To define such penalties, they need to derive from the \texttt{BasePenalty} class instead of the \texttt{SymmetricPenalty} one.
\elops then automatically adapts without further implementation requirements.
The only difference is that instead of specifying the functions \texttt{param\_slope}, \texttt{param\_limit}, and \texttt{param\_bndry}, users rather need to specify:
\begin{itemize}
    \item \texttt{param\_slope\_neg(self, i:int, lmbd:float)}: value of $\pertslope^-$ for some $\reg > 0$,
    \item \texttt{param\_slope\_pos(self, i:int, lmbd:float)}: value of $\pertslope^+$ for some $\reg > 0$,
    \item \texttt{param\_limit\_neg(self, i:int, lmbd:float)}: value of $\pertlimit^-$ for some $\reg > 0$,
    \item \texttt{param\_limit\_pos(self, i:int, lmbd:float)}: value of $\pertlimit^+$ for some $\reg > 0$,
    \item \texttt{param\_bndry\_neg(self, i:int, lmbd:float)}: value of $\pertrightlope^-$ for some $\reg > 0$,
    \item \texttt{param\_bndry\_pos(self, i:int, lmbd:float)}: value of $\pertrightlope^+$ for some $\reg > 0$,
\end{itemize}
where 
\begin{align}
    \pertslope^- &= \inf \kset{\bvi{}\in\kR-}{\conj{\pfunc}(\bvi{}) \leq \reg} \\
    \pertslope^+ &= \sup \kset{\bvi{}\in\kR+}{\conj{\pfunc}(\bvi{}) \leq \reg}
    \\
    \pertlimit^- &= 
    \begin{cases}
        \inf \kset{\bvi{}\in\kR-}{\bvi{} \in \subdiff\conj{\pfunc}(\pertslope^-)} &\text{ if } \subdiff\conj{\pfunc}(\pertslope^-) \neq \emptyset \\
        +\infty &\text{ otherwise}
    \end{cases}
    \\
    \pertlimit^+ &= 
    \begin{cases}
        \sup \kset{\bvi{}\in\kR+}{\bvi{} \in \subdiff\conj{\pfunc}(\pertslope^+)} &\text{ if } \subdiff\conj{\pfunc}(\pertslope^+) \neq \emptyset \\
        +\infty &\text{ otherwise}
    \end{cases}
    \\
    \pertrightlope^- &=
    \begin{cases}
        \inf\kset{\bvi{}\in\kR-}{\bvi{} \in \subdiff\pfunc(\pertlimit^-)} & \mbox{if \(\pertlimit^->-\infty\)}\\
        -\infty &\mbox{otherwise.}
    \end{cases}
    \\
    \pertrightlope^+ &=
    \begin{cases}
        \sup\kset{\bvi{}\in\kR+}{\bvi{} \in \subdiff\pfunc(\pertlimit^+)} & \mbox{if \(\pertlimit^+<+\infty\)}\\
        +\infty &\mbox{otherwise.}
    \end{cases}
\end{align}
are the counterparts of the quantities $(\pertslope,\pertlimit,\pertrightlope)$ defined in \eqref{eq:pertslope}-\eqref{eq:pertlimit}-\eqref{eq:pertrightlope} for non-symmetric penalties.

\subsection{Compiled Loss and Penalty Functions}
\label{sec:instances:compilation}

Custom loss and penalty functions defined by the user can derive from the \texttt{CompilableClass} template to take advantage of just-in-time compilation provided by \numba \citep{lam2015numba}.
This requires that all operations performed by the class are compatible with \numba.
Moreover, users additionally need to instantiate the two following methods:
\begin{itemize}
    \item \texttt{get\_spec(self)}: returns a Python \texttt{tuple} where each element corresponds to a \texttt{tuple} referring to one of the class attribute defined in the constructor of the user-defined function and specifying its name as a Python \texttt{str} as well as its \numba type.
    \item \texttt{params\_to\_dict(self)}: returns a Python \texttt{dict} where each key-value association refers to one of the class attribute defined in the constructor of the user-defined function and specifies its name as a Python \texttt{str} as well as its value.
\end{itemize}
This functionality is illustrated in \Cref{fig:custom-penalty} and is inspired from a similar one implemented in the \skglm package proposed by \cite{bertrand2022beyond}.

\section{Supplementary Material to the Numerical Experiments}
\label{sec:supp_numerics}

The code used for our experiments is open-sourced at
\begin{center}
    \url{https://github.com/TheoGuyard/l0exp}
\end{center}        
and all the datasets used are publicly available. Computations were carried out using the Grid’5000 testbed, supported by a scientific interest group hosted by Inria and including CNRS, RENATER and several universities as well as other organizations. Experiments were run on a Debian 10 operating system, featuring one Intel Xeon E5-2660 v3 CPU clocked at 2.60GHz with 16GB of RAM.

To calibrate the hyperparameters $(\reg, \regone, \regtwo, \bigM)$ in \Cref{sec:comparison_with_existing_toolboxes}, we use a grid search procedure.
For each dataset with $\dic \in \kR^{\ddim \times \pdim}$ and $\obs \in \kR^{\ddim}$, we first compute a solution
\begin{equation}
    \hat{\pv} \in \argmin_{\pv \in \kR^{\pdim}} \lfunc(\dic \pv)
\end{equation}
of the unregularized version of problem~\eqref{prob:prob}, with the appropriate loss function $\lfunc$ as defined in \Cref{tab:numerics}.
We also compute a value $\reg_{\max} > 0$ such that $\pv = \0$ is a solution of problem~\eqref{prob:prob} when $\reg = \reg_{\max}$.
Then, we explore all combinations of the following parameter ranges:
\begin{itemize}
    \item $\reg$: 10 logarithmically spaced values between $\reg_{\max}$ and $10^{-2} \reg_{\max}$,
    \item $\regone$: 7 logarithmically spaced values between $10^{-3} \ddim$ and $10^{3} \ddim$,
    \item $\regtwo$: 7 logarithmically spaced values between $10^{-3} \ddim$ and $10^{3} \ddim$,
    \item $\bigM$: 4 logarithmically spaced values between $10^0 \norm{\hat{\pv}}{\infty}$ and $10^3 \norm{\hat{\pv}}{\infty}$.
\end{itemize}
Depending on the penalty under consideration, we may fix either $\regone = 0$ or $\regtwo = 0$, in which case the corresponding parameter is not varied.
For each hyperparameter combination, we solve problem~\eqref{prob:prob} optimally and obtain a solution $\opt{\pv} \in \kR^{\pdim}$.
We then evaluate its associated Bayesian Information Criterion (BIC), as introduced by \citet{schwarz1978estimating} and defined as
\begin{equation}
    \text{BIC}(\opt{\pv}) = 2 \ddim \lfunc(\dic \opt{\pv}) + \log(\ddim)\norm{\opt{\pv}}{0},
\end{equation}
which balances loss minimization with model complexity.
Among all tested combinations, we select the one that achieves the lowest BIC value.
The selected hyperparameters are reported in \Cref{tab:hyperparameters}.
Although the original \mimosa solver sets its own hyperparameters, we have modified it to align with the values selected by our grid search procedure for consistency with the other solvers.

\begin{table}[!ht]
    \centering
    \scriptsize
    \begin{tabular}{llll|cccc}
        \toprule
        \textbf{Dataset} & \textbf{Dim. $\ddim \times \pdim$} & \textbf{Loss function} & \textbf{Penalty} & $\reg$ & $\regone$ & $\regtwo$ & $\bigM$ \\
        \midrule 
        & & & $\regone=0$, $\regtwo = 0$ & 0.0401 & 0 & 0 & 0.1235 \\
        \texttt{Riboflavin} & $71 \times 4,088$ & Least-squares & $\regone = 0$, $\regtwo > 0$ & 0.0087 & 0 & 7.1000 & 0.1235 \\
        & & & $\regone > 0$, $\regtwo > 0$ & 0.0074 & 0.0710 & 7.1000 & 0.1235 \\
        \midrule
        & & & $\regone = 0$, $\regtwo = 0$ & 2.0886 & 0 & 0 & 1.7816 \\
        \texttt{Leukemia} & $38 \times 7,129$ & Logistic & $\regone = 0$, $\regtwo > 0$ & 0.1486 & 0 & 3.8000 & 0.1782 \\
        & & & $\regone > 0$, $\regtwo > 0$ & 0.1452 & 0.0380 & 3.8000 & 0.1782 \\
        \midrule 
        & & & $\regone = 0$, $\regtwo = 0$ & 3.6358 & 0 & 0 & 0.7728 \\
        \texttt{Arcene} & $100 \times 10,000$ & Squared-hinge & $\regone = 0$, $\regtwo > 0$ & 0.3337 & 0 & 10.0000 & 0.0773 \\
        & & & $\regone > 0$, $\regtwo > 0$ & 0.3299 & 0.1000 & 10.0000 & 0.0773 \\
        \bottomrule
    \end{tabular}
    \caption{Hyperparameters selected for each instance considered in \Cref{tab:numerics}.}
    \label{tab:hyperparameters}
\end{table}
\begin{figure}[!ht]
  \centering
  \begin{minipage}{\codewidth\linewidth}
\begin{lstlisting}
import numpy as np
from numpy.typing import NDArray
from el0ps.datafit import BaseDatafit


class Huber(BaseDatafit):
    """Huber loss defined as f(w) = sum_{j=1}^m fj(wj) where
    fj(wj) = 0.5 * |wj - yj|^2 if |wj - yj| <= d and fj(wj) = d *
    (|wj - yj| - 0.5 * d) otherwise, with y in R^m and d > 0."""

    def __init__(self, y: NDArray, d: float):
        self.y = y
        self.d = d

    def value(self, w: NDArray):
        z = np.abs(w - self.y)
        v = 0.
        for zj in z:
            if zj <= self.d:
                v += 0.5 * zj ** 2
            else:
                v += self.d * (zj - 0.5 * self.d)
        return v
    
    def conjugate(self, w: NDArray):
        if np.any(np.abs(w) > self.d):
            return np.inf
        else:
            return np.sum(w + self.y)

    def gradient(self, w: NDArray):
        z = w - self.y
        g = np.empty_like(w)
        for j, zj in enumerate(z):
            if np.abs(zj) <= self.d:
                g[j] = zj
            else:
                g[j] = self.d * np.sign(zj)
        return g
    
    def gradient_lipschitz_constant(self):
        return 2. * self.d
\end{lstlisting}
  \end{minipage}
  \caption{Example of user-defined loss function.}
  \label{fig:custom-loss}
\end{figure}
\begin{figure}[!ht]
  \centering
  \begin{minipage}{\codewidth\linewidth}
\begin{lstlisting}
import numpy as np
from numba import float64
from el0ps.penalty import SymmetricPenalty
from el0ps.compilation import CompilableClass

class Berhu(SymmetricPenalty, CompilableClass):
    """BerHu penalty function defined as h(x) = d * |x| if |x| <= 1
    and h(x) = d * (x^2 + 1) / 2 otherwise for some d > 0."""

    def __init__(self, d: float):
        self.d = d
    
    def get_spec(self):
        return (('d', float64),)

    def params_to_dict(self):
        return {'d': self.d}

    def value(self, i: int, x: float):
        if np.abs(x) <= 1.:
            return self.d * np.abs(x)
        else:
            return self.d * (x ** 2 + 1.) / 2.
    
    def conjugate(self, i: int, x: float):
        return 0.5 * np.maximum(0., x ** 2 - self.d ** 2) / self.d

    def prox(self, i: int, x: float, eta: float):
        if np.abs(x) <= eta * self.d + 1.:
            return np.sign(x) * max(np.abs(x) - eta * self.d, 0.)
        else:
            return x / (1. + eta * self.d)
    
    def subdiff(self, i: int, x: float):
        if x == 0.:
            return [-self.d, self.d]
        elif np.abs(x) <= 1.:
            return [self.d * np.sign(x), self.d * np.sign(x)]
        else:
            return [self.d * x, self.d * x]

    def conjugate_subdiff(self, i: int, x: float):
        if np.abs(x) < self.d:
            return [0., 0.]
        elif x == -self.d:
            return [-1., 0.]
        elif x == self.d:
            return [0., 1.]
        else:
            return [x / self.d, x / self.d]

    def param_slope(self, i: int, lmbd:float):
        return np.sqrt(self.d ** 2 + 2. * lmbd * self.d)
\end{lstlisting}
  \end{minipage}
  \caption{Example of a user-defined penalty function with specified parameter $\pertslope$ and made compatible with \numba compilation.}
  \label{fig:custom-penalty}
\end{figure}
\clearpage

\end{document}